\documentstyle[aps,twocolumn,epsf,psfig]{revtex}

\begin{document}
\draft

\title {Franck-Condon Factors as Spectral Probes of Polaron Structure}

\author{
David W. Brown${^1}$,
Aldo~H.~Romero${^{2}}$, and
 Katja Lindenberg${^{1,3}}$
}

\address
{${^1}$
Institute for Nonlinear Science,\\
University of California, San Diego, La Jolla, CA 92093-0402}

\address
{${^2}$
Max-Planck Institut f\"{u}r Fest\-k\"{o}rper\-forschung,\\
Heisenbergstr. 1, 70569 Stuttgart, Germany}

\address
{${^3}$
Department of Chemistry and Biochemistry,\\
University of California, San Diego, La Jolla, CA 92093-0340} 

\date{\today} 

\maketitle

\begin{abstract}

We apply the Merrifield variational method to the Holstein molecular
crystal model in $D$ dimensions to compute non-adiabatic polaron band
energies and Franck-Condon factors at general crystal momenta.
We analyze these observable properties to extract characteristic
features related to polaron self-trapping and potential experimental
signatures.  These results are combined with others obtained by the
Global-Local variational method in 1D to construct a polaron phase
diagram encompassing all degrees of adiabaticity and all
electron-phonon coupling strengths.  The polaron phase diagram so
constructed includes disjoint regimes occupied by {\it small}
polarons, {\it large} polarons, and a newly-defined class of
{\it compact} polarons, all mutually separated by an intermediate
regime occupied by transitional structures.

\end{abstract}

\pacs{PACS numbers: 71.38.+i, 71.15.-m, 71.35.Aa, 72.90.+y}

\narrowtext

\section{Introduction}

The theory of polarons has undergone an evolution in recent years that
has substantially improved our ability to put solid, quantitative
accuracy to matters that have heretofore enjoyed only semi-quantitative
estimation or qualitative characterization.
This can be said in view of a convergence of results \cite{Romero98a}
that has been found in a number of independent and high-quality
methods that have been brought to bear in particular on the analysis
of the Holstein molecular crystal model \cite{Holstein59a,Holstein59b}.
Important among these methods are cluster
diagonalization \cite{Capone97,Wellein97a,deMello97,Alexandrov94a},
density matrix renormalization group \cite{Jeckelmann98a},
quantum Monte Carlo \cite{McKenzie96,DeRaedt83,DeRaedt84,Lagendijk85,Kornilovitch98a,Alexandrov98a},
and certain variational approaches \cite{Romero98a,Romero98g,Brown97b,Brown97a,Zhao97a,Zhao97b,Romero98c,Romero98e,Romero99b,Romero99c,Trugman99,Bonca98}.
Though quite distinct in their conception and implementation, these
methods have all been found to be in deep and broad quantitative
agreement over wide regions of the polaron parameter space.

Our own effort in this area has relied mainly upon the Global-Local
variational method, certain results of which will figure in the
present work.  A significant part of this effort has dealt with the
problem of developing a reliable and interpretable polaron phase
diagram on which can be clearly delineated the distinct regions of the
polaron parameter space where distinct classes of polaron structure
may be found.  In the course of this development, some familiar
notions that have become part of the polaron lore have had to be
revised, including the real physical nature of the large
polaron \cite{Romero99b,Romero99c,Romero98d} and the meaning of
self-trapping in $D$ dimensions \cite{Romero98f,Romero99a}.

Here, we continue to be concerned with the polaron phase diagram,
but in a manner and regime that are complementary to what has been
already been developed.  For practical and formal reasons, the utility
of the Global-Local variational method deteriorates significantly when
the fundamental electron transfer integrals are small and the
electron-phonon coupling is weak; as a practical matter, this
limitation excludes a sufficient portion of the non-adiabatic
regime to preclude a meaningful assessment of the self-trapping
transition there.  Necessarily, therefore, what we have been able
to say about the polaron phase diagram in the non-adiabatic regime
has been based on extrapolations from more adiabatic behaviors.

The non-adiabatic regime is important to many narrow-band systems
and particularly to molecular crystals for which the Holstein model
was originally formulated \cite{Holstein59a,Holstein59b,Kuper62,Kenkre82,Ueta86,Silinsh94,Song96}.
Polaron properties in the non-adiabatic regime generally depend
quite smoothly on the basic system parameters, without the relative
abruptness that tends to emerge in the adiabatic regime, and the
low orders of perturbation theory, either weak-coupling or
strong-coupling, tend to do a reasonable job of capturing most behaviors.
Paradoxically, perhaps, it is this relative unremarkableness of
the non-adiabatic regime that raises some of the questions motivating
our study; in particular, how the dramatic character of the
self-trapping transition that is so obvious at high adiabaticity
dissembles into relative obscurity, and how, as a practical matter,
its lingering presence may be recognized in observable polaron properties.

We approach this problem through the use of the Merrifield variational
method \cite{Zhao97a,Merrifield64}.
The Merrifield method can be viewed as an antecedent to the Global-Local
method in that it is the first in a sequence of increasingly refined
variational methods leading to the Global-Local method.
Although the Merrifield method suffers some very characteristic
limitations that restrict its usefulness as a tool for implementing
polaron theory at general points in the polaron parameter space, it is
at its best in the non-adiabatic regime where computation by the more
general Global-Local method becomes difficult, and is thus well-suited
to the present task. 
Moreover, owing to its relative simplicity, it is possible to pursue
results in $D$ dimensions, and to obtain some degree of analytical
guidance and support for numerical studies.

In order to locate the self-trapping transition, we need to
analyze observable physical properties that take on distinguishable
asymptotic behaviors on each side of the transition and objectively
locate a boundary that discriminates between these behaviors.
Here, we focus on two properties that are particularly important
to spectral studies, the polaron ground state energy (related in
well-known ways to Stokes shifts) and Franck-Condon factors (related
in well-known ways to oscillator strengths).
As a function of the electron-phonon coupling strength $g$, the
ground state energy generally exhibits a ``knee'' between distinct
weak- and strong-coupling trends that can be located and followed
in parameter space to develop a self-trapping line.
Franck-Condon factors generally exhibit a distorted Gaussian dependence
on the coupling strength, allowing the central peak region (weak coupling)
to be objectively distinguished from the outer tail region (strong coupling).

We use the Holstein Hamiltonian \cite{Holstein59a,Holstein59b} on
a $D$-dimensional Euclidean lattice
\begin{eqnarray}
\hat{H} &=& - \sum_{ \vec{n} } \sum_{i=1}^D
J_i a_{\vec{n}}^{\dagger}
( a_{\vec{n}+\vec{\epsilon}_i} + a_{\vec{n}-\vec{\epsilon}_i} )
 \nonumber \\
&& + \hbar \omega \sum_{\vec{n}} b_{\vec{n}}^{\dagger} b_{\vec{n}}
- g \hbar \omega \sum_{\vec{n}} a_{\vec{n}}^{\dagger}
a_{\vec{n}} ( b_{\vec{n}}^{\dagger} + b_{\vec{n}} ) ~,
\end{eqnarray}
in which $a_{\vec{n}}^\dagger$ creates a single electronic excitation
in the rigid-lattice Wannier state at site ${\vec{n}}$,
and $b_{\vec{n}}^\dagger$ creates a quantum of vibrational
energy $\hbar\omega$  in the Einstein oscillator at site ${\vec{n}}$.
The $J_i$ are the nearest-neighbor electronic transfer integrals
along the primitive crystal axes, and the $\hat{\epsilon}_i$ are
unit vectors associated with the primitive translations.
The above model encompasses all Bravais lattices, with the different
lattice structures appearing only in the relative values of the
hopping integrals $J_i$.
For simplicity in the following, we use terms appropriate to
orthorhombic lattices in which conventionally $i = x $, $y$, or $z$;
however, all results hold for lattices of lower symmetry with
appropriate transcription of these labels to those of the primitive axes.

We use the following Fourier conventions for ladder operators
($c^{\dag} = a^{\dag} , b^{\dag}$) and scalars:
\begin{equation}
c^{\dag}_{\vec{n}}  = N^{-D/2} \sum_{\vec{p}} e^{-i{\vec{p}}\cdot{\vec{n}}} c^{\dag}_{\vec{p}}, ~~~
c^{\dag}_{\vec{p}}  = N^{-D/2} \sum_{\vec{n}} e^{i{\vec{p}}\cdot{\vec{n}}} c^{\dag}_{\vec{n}},
\end{equation}
\begin{equation}
\lambda_{\vec{n}}^{\vec{\kappa}}~=~ N^{-D} \sum_{\vec{q}} e^{i{\vec{q}}\cdot{\vec{n}}} \lambda_{\vec{q}}^{\vec{\kappa}}, ~~~~
\lambda_{\vec{q}}^{\vec{\kappa}}~=~ \sum_{\vec{n}} e^{-i{\vec{q}}\cdot{\vec{n}}} \lambda_{\vec{n}}^{\vec{\kappa}} ~.
\label{eq:fourier}
\end{equation}

It is convenient in the following to characterize tunneling
strength in $D$ dimensions in part through a parameter
${\cal{J}} = \sum_i J_i$; when restricting discussion to
isotropic tunneling, we use the notation $J = J_i$, such that
${\cal{J}} = DJ$ in those cases.

For the most part in this paper, we limit our discussion to the
non-adiabatic regime, defined by the condition ${\cal{J}}/\hbar\omega < 1/4$.
Polarons at such small ${\cal{J}}/\hbar\omega$ are quite narrow
since we know that the {\it largest} polaron in any dimension (as
characterized by the size of the phonon cloud) has a width of
$\sqrt{2 J_i /\hbar \omega}$ along the $i$ axis \cite{Romero98d};
since no $J_i /\hbar\omega$ is greater than $1/4$ in the non-adiabatic
regime, no polaron in this regime has a width greater than a lattice
constant, even at vanishing coupling.
Thus, the variational space in which the problem is solved numerically
need {\it not} be large in order to contain the complete polaron.
This ability to contain the present problem in a small real-space
volume facilitates computation considerably.

The Merrifield trial state may be written
\begin{eqnarray}
| \Psi ( \vec{\kappa} ) \rangle &=& N^{-D/2} \sum_{\vec{n}} e^{i \vec{\kappa} \cdot \vec{n}} a^{\dag}_{\vec{n}}
\nonumber \\
&& \times \exp [ - N^{-D/2} \sum_{\vec{q}} ( \lambda^{\vec{\kappa}}_{\vec{q}} e^{-i{\vec{q}}\cdot{\vec{n}}}
 b^{\dag}_{\vec{q}} - h.c. )] |0\rangle ~,
\\
\langle \Psi (\vec{\kappa}) | & \Psi & (\vec{\kappa} ') \rangle = \delta_{\vec{\kappa} \vec{\kappa}'} ~,
\end{eqnarray}
in which the $\{ \lambda_{\vec{q}}^{\vec{\kappa}}\}$ are the
variational parameters specifying the coherent state amplitude in
the phonon mode $\vec{q}$.
Though these trial states are delocalized and satisfy the appropriate
Bloch symmetry condition, and thus any property measured in the ``lab''
frame is uniform over the lattice, the internal structure of these
delocalized states is determined by exciton-phonon correlations that
are essentially local in character.
Here, that local character is such that the electronic component
located at site $\vec{n}$ is associated with a ``phonon cloud''
centered on that site, determined by the set of lattice amplitudes
$\{ \lambda_{\vec{q}}^{\vec{\kappa}}\}$.

We evaluate the variational energy band as the expectation value
of the Holstein Hamiltonian
\begin{eqnarray}
\label{h1}
E^{\vec{\kappa}} &=& \langle  \Psi ( \vec{\kappa} ) | \hat{H} |\Psi ( \vec{\kappa} ) \rangle  = - \sum_{i=1}^D J_i ( e^{i \kappa_i } S^{\vec{\kappa} *}_{+i} + e^{-i \kappa_i} S^{\vec{\kappa}}_{-i} ) 
\nonumber \\
&& + N^{-D} \hbar \omega \sum_{\vec{q}} | \lambda^{\vec{\kappa}}_{\vec{q}} |^2 
- N^{-D} g \hbar \omega \sum_{\vec{q}} ( \lambda^{\vec{\kappa} *}_{-\vec{q}} + \lambda^{\vec{\kappa}}_{\vec{q}} ),
\end{eqnarray}
wherein $S^{\vec{\kappa}}_{\pm i}$ is the Debye-Waller factor
\begin{equation}
S^{\vec{\kappa}}_{\pm i}  = \exp [ N^{-D} \sum_{\vec{q}} | \lambda^{\vec{\kappa}}_{\vec{q}} |^2 ( e^{\pm i q_i } - 1 ) ] .
\end{equation}

Minimization of $ E^{\vec{\kappa}} $ with respect to
$ \lambda^{\vec{\kappa} *}_{\vec{q}}$ leads to the self-consistency
equations for $ \lambda_{\vec{q}}^{\vec{\kappa}} $:
\begin{equation}
\label{eq:lambda}
\lambda^{\vec{\kappa}}_{\vec{q}}  = \frac { g \hbar \omega}
{
 \hbar \omega - \sum_{i=1}^D \left[ 4 J_i S^{\vec{\kappa}}_{i} \sin (\kappa_i- \Phi^{\vec{\kappa}}_i  - \frac {q_i} 2 ) \sin \frac {q_i} 2 \right]
} ,
\end{equation}

\begin{equation}
\label{eq:Sk}
S^{\vec{\kappa}}_i  = \exp [ N^{-D} \sum_{\vec{q}} | \lambda_{\vec{q}}^{\vec{\kappa}} |^2 (\cos q_i -1) ] ~,
\end{equation}
\begin{equation}
\label{eq:phik}
\Phi^{\vec{\kappa}}_i  = N^{-D} \sum_{\vec{q}} | \lambda_{\vec{q}}^{\vec{\kappa}} |^2 \sin q_i ~,
\end{equation}
\begin{equation}
S_i^{\vec{\kappa}} = S_i^{- \vec{\kappa}}~, ~~~ \Phi_i^{\vec{\kappa}} = - \Phi_i^{-\vec{\kappa}} ~,~~~ \lambda_{\vec{q}}^{\vec{\kappa}} = \lambda_{-\vec{q}}^{-\vec{\kappa}} ~,
\label{eq:symm}
\end{equation}
where $S^{\vec{\kappa}}_i$ and $\pm \Phi^{\vec{\kappa}}_i$ are the
magnitudes and phases of the complex Debye-Waller factors
$S_{\pm i}^{\vec{\kappa}}$.
This shows the optimal $\lambda_{\vec{q}}^{\vec{\kappa}}$ to be real,
and establishes the ``sum rule''
\begin{equation}
\sum_{\vec{n}} \lambda_{\vec{n}}^{\vec{\kappa}} = \lambda_{{\vec{q}}=0}^{\vec{\kappa}} = g ~,
\label{eq:sumrule}
\end{equation}
valid at any $\vec{\kappa}$ and in any number of dimensions.

When any particular $J_i/\hbar\omega \rightarrow 0$,
$\lambda_{\vec{q}}^{\vec{\kappa}}$ loses any dependence on $q_i$
and $\kappa_i$, becoming ``flat'' in those variables.
The Debye-Waller factor ($S_i^{\vec{\kappa}}$) and phase
($\Phi_i^{\vec{\kappa}}$) associated with that direction drop out of
the problem and the real space phonon amplitudes
$\lambda_{\vec{n}}^{\vec{\kappa}}$ become completely localized
along the $i$ axis.
Although a disparity among the relative magnitudes of several $J_i$
can result in a polaron that is in respects ``small'' in certain
directions and ``large'' in others, there is not a distinct
self-trapping transition associated with each $J_i$ \cite{Romero99a}.
This can be seen in the present context in the fact that the
Debye-Waller factor $S_z^{\vec{\kappa}}$ associated with a vanishing
$J_z$ does not approach $e^{-g^2}$, which would be expected of 1D small
polarons along the $z$ axis, but a quantity that is characteristic of
the 2D polaron structure in the two surviving dimensions, whether this
be large-polaron-like or small-polaron-like.

The sum rule continues to be satisfied as dimensions are turned
off or on (e.g., $J_z \rightarrow 0$ in three dimensions), since
\begin{eqnarray}
\lim_{J_z \rightarrow 0} && \sum_{n_x,n_y,n_z} \lambda_{(n_x,n_y,n_z)}^{(\kappa_x,\kappa_y,\kappa_z)} =
\nonumber \\
&& \sum_{n_x,n_y,n_z} \lambda_{(n_x,n_y)}^{(\kappa_x,\kappa_y)} \delta_{n_z , 0} =
\sum_{n_x,n_y} \lambda_{(n_x,n_y)}^{(\kappa_x,\kappa_y)} = g ~.
\end{eqnarray}
Thus, there is no need for dimension-specific formulation if
dimensions are controlled through the tuning of $\{ J_i/\hbar\omega \}$.

Owing to the symmetries (\ref{eq:symm}) and the periodicity of the
reciprocal lattice, $\Phi_i^{\vec{\kappa}}$ vanishes at the Brillouin
zone center and everywhere on the Brillouin zone boundary.
Consequently, we have certain special values that play a significant
role in the following.
Denoting the reciprocal lattice origin by $\vec{0}$ and any of the
most extreme points of the Brillouin zone by
$\vec{\pi}$ ($\kappa_i = \pm \pi$ along each axis), we find 
\begin{equation}
\lambda^{\vec{0}}_{\vec{q}}  = \frac { g \hbar \omega}
{
 \hbar \omega + \sum_{i=1}^D \left[ 4 J_i S^{\vec{0}}_{i} \sin^2 \frac {q_i} 2 \right]
} ~,
\label{eq:lambda0}
\end{equation}

\begin{equation}
\lambda^{\vec{\pi}}_{\vec{q}}  = \frac { g \hbar \omega}
{
 \hbar \omega - \sum_{i=1}^D \left[ 4 J_i S^{\vec{\pi}}_{i} \sin^2 \frac {q_i} 2 \right]
} ~.
\label{eq:lambdapi}
\end{equation}

The zone-center phonon amplitudes (\ref{eq:lambda0}) are well-behaved
under all circumstances because the denominators, similar to those of
weak-coupling perturbation theory, are sums of bounded positive terms.
The zone-edge relation (\ref{eq:lambdapi}), on the other hand,
suggests the possibility of encountering large or divergent phonon
amplitudes for phonon wave vectors near the Brillouin zone boundary
if tunneling is sufficiently strong (${\cal{J}}/\hbar\omega \ge 1/4$)
and the Debye-Waller factors $\{S_i^{\vec{\pi}} \}$ are sufficiently
near unity (as generally occurs when electron-phonon coupling is
sufficiently weak; see the 1D examples below).
This potential divergence is both a real physical phenomenon and
a generator of artifacts in the Merrifield method.

The reality of the phenomenon is due to the resonance or near-resonance
that can occur between the states of the one-phonon continuum and
zone-edge states of both the free-electron and the self-consistent
polaron when the energy gap between latter and the one-phonon continuum
is small.  This circumstance occurs in any number of dimensions when
${\cal{J}}/\hbar\omega \gtrsim 1/4$ and $g$ is small.
Under these circumstances, only a very small amount of electron-phonon
coupling is needed to produce intense interactions that flatten the
outer polaron energy band (level repulsion) and create heavy phonon
clouds strongly modulated by the character of the zone-edge phonons.

The Merrifield method accommodates the nearness of the one-phonon
continuum by producing strong distortions of the variational
amplitudes of a qualitatively appropriate nature; the phonon
amplitudes become highly focussed around a single phonon wave
vector, in clear approximation to the single-phonon quantum that
constitutes the exact $g \rightarrow 0$ state.
However, because the Merrifield state is not well-equipped to
emulate the highly quantum mechanical character of such states, the
energy balance central to the variational principle is distorted and
the variational energy bound is raised.
Consequently, rather than experiencing the expected strong repulsion
from the one-phonon continuum, the outer-zone portions of Merrifield
energy bands flatten relatively weakly and cannot be taken as
appropriately indicative of the polaron structure when
${\cal{J}} /\hbar\omega \gtrsim 1/4$.

Although one cannot rely upon the numerical values of Merrifield
band energies influenced by resonances with the one-phonon continuum,
it is nonetheless true that the general character of the variational
lattice state responds to such resonance effects in a reasonably
appropriate way, provided that ${\cal{J}}/\hbar\omega \le 1/4$.
This suggests that changes in the variationally-determined Franck-Condon
factors, as very direct figures of merit for this general character,
may reasonably indicate where the essential changes in polaron
structure occur.
It is thus that in the following we rely upon Franck-Condon factors
as our primary diagnostic of outer-zone polaron structure.
Although this proves to be a very practical election, it is a choice
that is in respects forced upon us by the limitations of the
Merrifield method.
As a choice that in the larger picture should be sufficient, but
not necessary, other theoretical methods not so limited should
find similar behavior in the band energies and other polaron
properties near the band edge.

\section{The 1D case}

In principle, the set of equations (\ref{eq:lambda}) - (\ref{eq:phik})
can be closed in the $S_i^{\vec{\kappa}}$ and $\Phi_i^{\vec{\kappa}}$
alone, greatly reducing the size of the self-consistency problem to
be solved.  This is of practical advantage only in one dimension,
however, since the reduction to quadratures involved in higher
dimensions does not significantly facilitate computation.

\begin{figure}[htb]
\begin{center}
\leavevmode
\epsfxsize = 3.6in
\epsffile{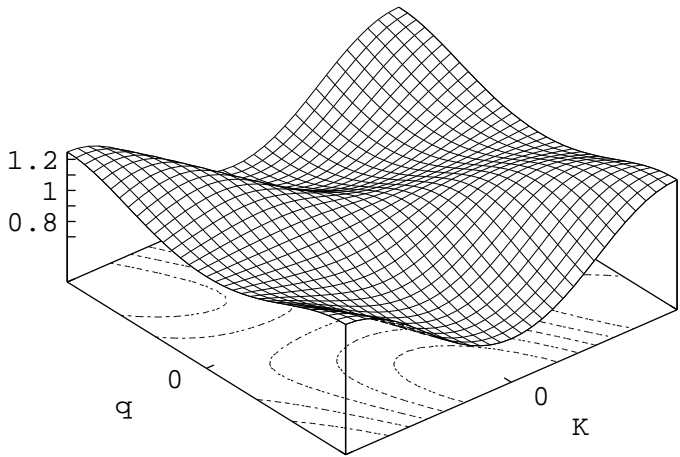}
\end{center}
\vspace{0.2in}
\caption
{
Sample surface showing the (real) variational amplitudes $\lambda_q^\kappa$
in the 1D case for $J/\hbar\omega = 0.2$ and $g=1$.
}
\label{fig:j.2g1}
\end{figure}

Replacing the summations in Eqs.~(\ref{eq:Sk}) and (\ref{eq:phik})
with 1D integrations, one arrives at the self-consistency equations
first obtained by Merrifield: 
\begin{equation}
\label{eq:s}
S^\kappa = \exp ( - g^2 \Delta^\kappa ) ~,
\end{equation}
\begin{equation}
\label{eq:phi}
\Phi^\kappa = - g^2 \Delta^\kappa (2JS^\kappa/\hbar\omega) \sin (\Phi^\kappa - \kappa)  ~,
\end{equation}
and 
\begin{equation}
\label{eq:dk}
\Delta^\kappa = [ ~ \frac {\hbar\omega} { \sqrt { [\hbar\omega+2JS^\kappa \cos(\Phi^\kappa -\kappa ) ]^2 -(2JS^\kappa)^2 } } ~ ]^3 ~.
\end{equation}
Using (\ref{eq:s})~-~(\ref{eq:dk}) in (\ref{eq:lambda}) yields the
full set of variational phonon amplitudes such as shown in
Figure~\ref{fig:j.2g1}.

One may further obtain the energy-momentum relation:
\begin{eqnarray}
E^\kappa &=& g^2 \hbar\omega (\Delta^\kappa-2{[\Delta^\kappa]} ^{\frac 1 3})
\nonumber \\
&& -2JS^\kappa(1-g^2 \Delta^\kappa) \cos (\kappa-\Phi^\kappa) .
\label{eq:Ek}
\end{eqnarray}
The price paid for the compactness of this expression is the
self-consistency condition that makes (\ref{eq:Ek}) awkward to
analyze; however, it can be shown that (\ref{eq:Ek}) agrees with
weak-coupling perturbation theory through second order in $g$, and
strong-coupling perturbation theory through first order in $J/\hbar\omega$.
The latter is actually a shortcoming of the Merrifield method since
important contributions from second order quickly dominate the first
order of strong-coupling perturbation theory; however, the first
order is sufficient to properly determine that in the
$J/\hbar\omega \rightarrow 0$ limit the ``knee'' in the dependence
of $E^{\kappa}$ on $g$ at any $\kappa$ is given by
\begin{equation}
\frac {\partial^3 E^{\kappa}} {\partial g^3} = 0 ~~\Rightarrow~~ g= \sqrt{\frac 3 2} ~~~ at ~~~ \frac J {\hbar\omega} = 0
\end{equation}
The same differential criterion can be applied at finite $J/\hbar\omega$,
which we will use to develop the phase diagram in Section IV below.

\section{The Franck-Condon factor}

A quantity intimately related to the Debye-Waller factors
appearing in the self-consistency equations is the Franck-Condon factor
\begin{equation}
F(\vec{\kappa}) = | \langle \Psi ( \vec{\kappa} ) | a_{\vec{\kappa}}^{\dagger} | 0 \rangle | ^2 ~.
\end{equation}
This is one of many Franck-Condon factors associated with various
transitions between correlated electron-phonon states.
This particular factor characterizes the oscillator strength of
the zero-phonon line associated with a transition between a free
electron state of crystal momentum $\vec{\kappa}$ and the polaron it
forms at the same $\vec{\kappa}$; these are direct transitions,
resolved by crystal momentum.
Though one is often concerned primarily with transitions near the
Brillouin zone center, we will use the full $\vec{\kappa}$ dependence
of $F(\vec{\kappa})$ across the Brillouin zone, and particular at the
zone center and the zone boundary.

Our principal interest in zero-phonon lines in the present context is
in the possibility that they may offer an observable means for
mapping the essential polaron features on the polaron phase diagram.
Owing to the strong similarity between the Franck-Condon factor
and the Debye-Waller factors intimately connected with the polaron
effective mass, it is reasonable to expect that an analysis of the
dependence of the Franck-Condon factor on model parameters should be
able to yield the location of the self-trapping line.
Also owing to the fact that the Merrifield method is at its best in
the non-adiabatic regime and weak coupling, it is reasonable to hope
that such an analysis would complement others made by other methods
generally more accurate (e.g., the Global-Local method), but which
deteriorate in quality at small ${\cal{J}}/\hbar\omega$ and $g$.

The quenching of the zero-phonon line is an experimental signature
that has long been associated with the self-trapping transition.
As a function of model parameters, this quenching occurs continuously
as electron-phonon coupling is tuned from the weak-coupling to the
strong-coupling regimes.
There is thus some inherent ambiguity in the assignment of the point
we associate with the self-trapping transition; our criterion here,
as elsewhere, is to identify the self-trapping transition as the
point of {\it most rapid change} in a property that takes on
characteristically different behaviors in the weak and strong-coupling
regimes.  In the case of the Franck-Condon factor, this criterion
takes the form of an inflection point in the dependence of $F(\vec{\kappa})$
on $g$ at fixed $\{ J_i \}$.

In terms specific to our variational development,
\begin{equation}
F( \vec{\kappa} ) = \exp \left( - N^{-D} \sum_{\vec{q}} | \lambda_{\vec{q}}^{\vec{\kappa}} |^2 \right) ~,
\end{equation}
which is just the exponential of the average number of phonons per
mode in the phonon cloud.

\begin{figure}[htb]
\begin{center}
\leavevmode
\epsfxsize = 3.6in
\epsffile{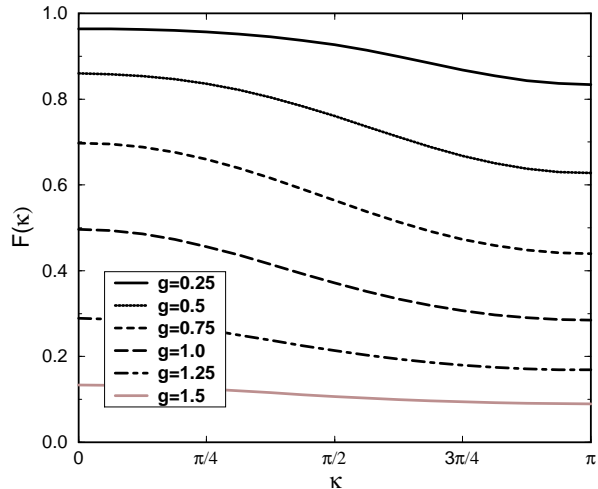}
\end{center}
\caption
{
Franck-Condon factors across the Brillouin zone in 1D.
$J/\hbar\omega = 0.2$, $g = 0.25 - 1.5$.
}
\label{fig:fofk1d}
\end{figure}

In the 1D case, we find
\begin{equation}
F(\kappa) = \exp \left\{ - {g^2} \Delta^{\kappa} [ 1-(2JS^{\kappa}/\hbar\omega) \cos (\kappa - \Phi^{\kappa} ) ] \right\} ~,
\label{eq:fc1dgeneral}
\end{equation}
as shown in Figure~\ref{fig:fofk1d} in selected cases.

\begin{figure}[htb]
\begin{center}
\leavevmode
\epsfxsize = 3.6in
\epsffile{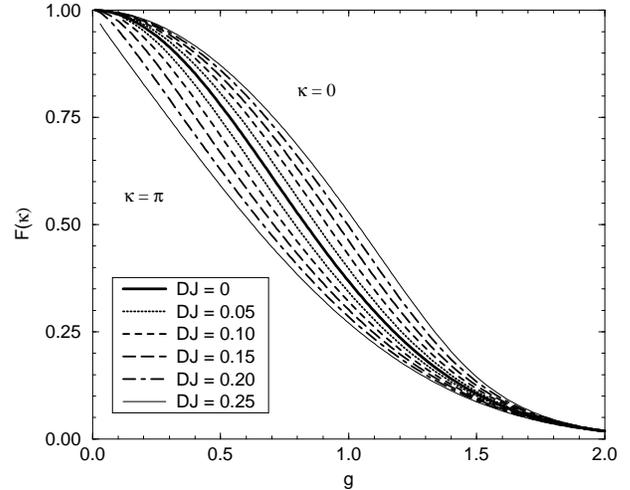}
\end{center}
\caption
{
Franck-Condon factors in $D$ dimensions at the zone center (upper curves)
and at the most extreme point of the Brillouin zone ($\kappa_i = \pi$)
for $DJ/\hbar\omega = 0~-~0.25$.
}
\label{fig:fofk0pi}
\end{figure}

It is simple to show that in the absence of tunneling
$F(\vec{\kappa}) = e^{-g^2}$ for all $\vec{\kappa}$ in any number
of dimensions.
Thus, in the ${\cal{J}}/\hbar\omega \rightarrow 0$ limit the
self-trapping features associated with the Franck-Condon factor
are found at $g= 1/\sqrt{2}$, for any $\vec{\kappa}$.
With increasing tunneling, this degeneracy is broken and these
features fan out;
the manner in which this occurs can be seen most clearly in the 1D
case, where
\begin{eqnarray}
F(\kappa=0) &=& \exp \left\{ 
- g^2 \frac {1-2JS^0/\hbar\omega} {(1+4JS^0 /\hbar\omega )^{3/2}}
\right\} ~,
\label{eq:fc1dk0} \\
F(\kappa=\pi) &=& \exp \left\{ 
- g^2 \frac {1+2JS^\pi/\hbar\omega} {(1-4JS^\pi /\hbar\omega )^{3/2}}
\right\} ~.
\label{eq:fc1dkpi}
\end{eqnarray}
(See Figure \ref{fig:fofk0pi}.)
It is clear from these that the Franck-Condon factor is independent
of $\kappa$ at $J/\hbar\omega =0$.
It is also clear that increasing $J/\hbar\omega $ from $0$ causes
$F(0)$ to broaden out to stronger coupling and causes $F(\pi)$ to
narrow toward weaker coupling.

\begin{figure}[htb]
\begin{center}
\leavevmode
\epsfxsize = 3.6in
\epsffile{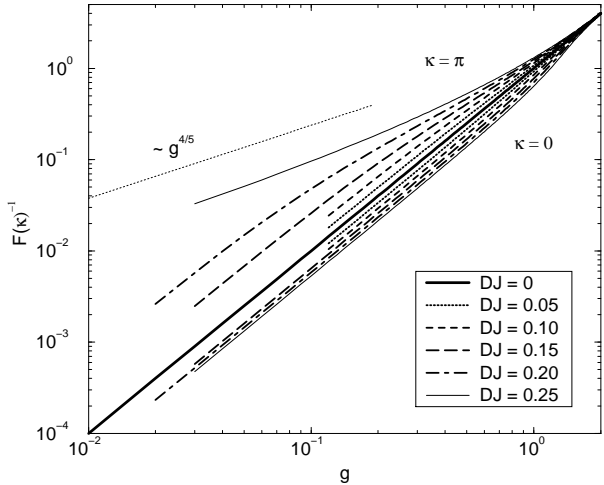}
\end{center}
\caption
{
$\log (-\ln F(\vec{\kappa}))$ vs. $\log g$, allowing the power of $g$ in
the exponent of the Franck-Condon factor to be ascertained.
Curves are coded to correspond to Figure~\ref{fig:fofk0pi}; zone
center (lower curves) and zone edge (upper curves) in $D$ dimensions
for $DJ/\hbar\omega = 0$, $0.05$, $0.1$, $0.15$, $0.2$, and $0.25$.
}
\label{fig:fofk0pilog.eps}
\end{figure}

Perhaps the most interesting behavior revealed in
(\ref{eq:fc1dgeneral})~-~(\ref{eq:fc1dkpi}) is that of $F(\pi)$ at
$J/\hbar\omega=1/4$.
Using the fact that at this particular $J/\hbar\omega$ value
\begin{equation}
S^{\pi} = \exp [ - g^2 (1- S^{\pi})^{-3/2} ] ~,
\end{equation}
one can show that the leading dependence of $S^{\pi}$ on $g$ is
\begin{equation}
S^{\pi} \sim 1 - g^{4/5} ~.
\end{equation}
This in turn implies that in the same approximation,
\begin{equation}
F(\pi) \sim \exp [ - \frac 3 2 g^{4/5} ] ~.
\end{equation}
(See Figure \ref{fig:fofk0pilog.eps}.)
This singular behavior in the Franck-Condon factor at
$J/\hbar\omega = 1/4$ suggests that the self-trapping feature identified
by an inflection point in $g$ moves to $g=0$ at $J/\hbar\omega = 1/4$.
Indeed, on can show that for $J/\hbar\omega = 1/4 - \epsilon$, the
Franck-Condon factor retains an initial finite negative curvature in
$g$, suggesting that there exists a proper inflection point at finite $g$.

Thus, all these considerations suggest that the self-trapping feature
in $F(0)$ should should shift more-or-less steadily from
$g= 1/ \sqrt{2}$ at $J/\hbar\omega=0$ to {\it stronger} coupling with
increasing $J/\hbar\omega$, while the self-trapping feature in $F(\pi)$
appears to shift from the same $g= 1/\sqrt{2}$ to {\it weaker} coupling
with increasing $J/\hbar\omega$, arriving at $g=0$ at $J/\hbar\omega= 1/4$.

The particular results above have been obtained for the 1D case, which
enjoys sufficient tractibility to admit some reasonably straightforward
formal analysis.
In higher dimensions, numerical solution and analysis is generally
more practical, though certain exceptions warrant special attention.
Throughout the foregoing we have highlighted the particular crystal
momentum values associated with the Brillouin zone center
($\vec{\kappa} = \vec{0}$) and the most remote corners of the Brillouin
zone where all the crystal momentum components take their maximum
values ($\vec{\kappa} = \vec{\pi}$).
We demonstrate in Appendix A that for these particular $\vec{\kappa}$
values in the isotropic cases in $D$ dimensions, the dependence of the
Franck-Condon factors on the dimensionality and tunneling strength is
reduced to the single scaled variable $DJ$.
This implies that the particular results shown in Figure~\ref{fig:fofk0pi}
hold not only in 1D, but in 2D and 3D as well when parameters are
scaled appropriately.

Results can be obtained numerically for any degree of anisotropy and
general $\vec{\kappa}$; however, the qualitative nature of the
dependence on anisotropy can be inferred from Figure~\ref{fig:fofk0pi}
without detailed analysis.
Consider, for example, $\vec{\kappa}~=~\vec{0}$ and $J/\hbar\omega~=~0.05$:
The results for quasi-2D scenarios with
$J_x/\hbar\omega~=~J_y/\hbar\omega~=~0.05$ and $0~<~J_z/\hbar\omega~<~0.05$
are contained between the $DJ/\hbar\omega~=~0.15$ and $DJ/\hbar\omega~=~0.1$
cases shown in Figure~\ref{fig:fofk0pi}.
Similarly, the results for quasi-1D scenarios with $J_x/\hbar\omega~=~0.05$,
$0~<~J_y/\hbar\omega~<~0.1$ and $J_z/\hbar\omega~=~0$ are contained
between the $DJ/\hbar\omega~=~0.05$ and $DJ/\hbar\omega~=~0.1$ cases
shown in Figure~\ref{fig:fofk0pi}.

The tunneling parameters and the effective dimensionality determined by
them are not generally subject to any practical degree of experimental
control; however, even greater changes in the Franck-Condon factor can
be induced by changing the magnitude and/or the orientation of the
crystal momentum $\vec{\kappa}$ probed.
To the extent that it is possible to achieve some selectivity in
the $\vec{\kappa}$'s sampled in a particular experiment, it should
be possible to induce controlled variations in the oscillator
strengths associated with these Franck-Condon factors by, for
example, varying the orientation of the sample.
Such predictable ``wobbles'' in the intensities of appropriately-selective
spectral probes constitute {\it signatures} of polaron structure that
exist only if {\it both} electron-phonon coupling {\it and} electron
tunneling are sufficiently great, {\it without} reducing the
quasiparticle to the status of a ``mere'' small polaron.

\section{Phase Diagram}

\begin{figure}[htb]
\begin{center}
\leavevmode
\epsfxsize = 3.6in
\epsffile{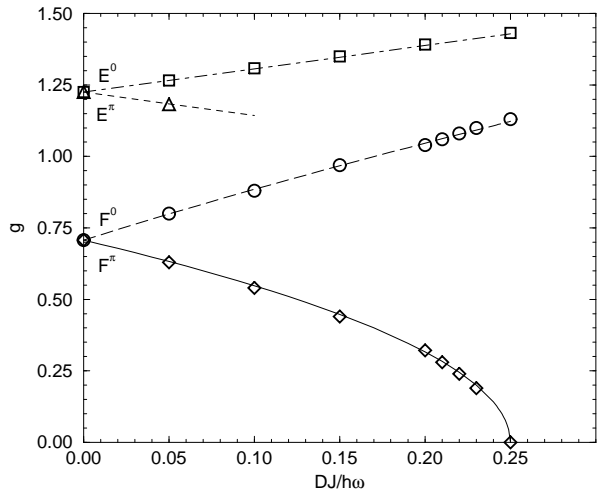}
\end{center}
\caption
{
Polaron phase diagram in the non-adiabatic regime according to the
Merrifield method in $D$ dimensions.
Diamonds:  Location of inflection points in $F(\pi)$.
Circles:  Location of inflection points in $F(0)$.
Squares:  Location of the knee in $E^0$.
Triangles:  Location of the knee in $E^{\pi}$.
Solid line, Eq.~\ref{eq:gfpi}.
Long-dashed line, Eq.~\ref{eq:gf0}.
Short-dashed line, Eq.~\ref{eq:gepi}.
Chain-dotted line, Eq.~\ref{eq:ge0}.
}
\label{fig:phase}
\end{figure}

The overall character of the foregoing results can be summarized on a
diagram of the polaron parameter space in which the loci of the knees
in the polaron band energies and the inflection points of Franck-Condon
factors play the role of rough phase boundaries separating distinct
classes of polaron structure (see Figure~\ref{fig:phase}).
These lines are accurately described by the simple relations
\begin{eqnarray}
g_{E^0} &\sim& \sqrt{ \frac 3 2} \left( 1 + \frac 2 3 \frac {DJ} {\hbar\omega} \right), ~~~~~ ~~ DJ/\hbar \omega \lesssim 1/4 ~,
\label{eq:ge0}
\\
g_{E^\pi} &\sim& \sqrt{ \frac 3 2} \left( 1 - \frac 2 3 \frac {DJ} {\hbar\omega} \right), ~~~~~ ~~ DJ/\hbar \omega \ll 1/4 ~,
\label{eq:gepi}
\\
g_{F^0} &\sim& \frac 1 {\sqrt{2}} \left( 1 + 4 \frac {DJ} {\hbar\omega} \right) ^ {2/3}, ~~~~~ DJ/\hbar \omega \lesssim 1/4 ~,
\label{eq:gf0}
\\
g_{F^\pi} &\sim& \frac 1 {\sqrt{2}} \left( 1 - 4 \frac {DJ} {\hbar\omega} \right) ^ {1/2}, ~~~~~ DJ/\hbar \omega < 1/4 ~.
\label{eq:gfpi}
\end{eqnarray}
obtained in empirical fashion by noting the exact ${\cal{J}} = 0$
termini as discussed in prior sections and augmenting these with the
simplest expressions in whole numbers that express the apparent
trends in a quantitatively consistent way.
The restrictions on (\ref{eq:ge0}) and (\ref{eq:gf0}) are weak because
such zone-center properties are well-behaved under the Merrifield
method to ${\cal{J}}/\hbar\omega$ substantially greater than $1/4$;
however, the quantitative accuracy of the Merrifield method even at
the zone center deteriorates significantly with increasing
${\cal{J}}/\hbar\omega$, warranting prudence beyond the strictly
non-adiabatic regime.
On the other hand, the restrictions on (\ref{eq:gepi}) and
(\ref{eq:gfpi}) are strong because zone-edge properties are
strongly affected by the one-phonon continuum.

These boundaries only roughly distinguish distinct polaron regimes
because the structural changes occurring in the non-adiabatic regime
are quite smooth and broad, with changes in different aspects of
polaron structure occurring with less synchronization than is seen
in the adiabatic regime.
This is seen clearly in the fact that the critical features of the
band energies and the Franck-Condon factors are significantly separated
in $g$ in the ${\cal{J}}/\hbar\omega \rightarrow 0$ limit, and while
trending similarly with increasing ${\cal{J}}/\hbar\omega$, remain
well separated over the entire non-adiabatic regime.

This disperse character of the collection of self-trapping-related
loci is not an artifact of the Merrifield method, nor of the
particular parameter regime nor of the particular physical quantities
used to locate transition effects.
A similar and complementary dispersion has been found at somewhat
larger $J/\hbar\omega$ in 1D using the Global-Local method, based
on the analysis of physical quantities such as the ground state
energy, kinetic energy, phonon energy, electron-phonon interaction
energy \cite{Romero98c}, effective mass \cite{Romero98e}, and
electron-phonon correlation functions \cite{Romero99b}.
In such analyses, the self-trapping loci attributable to different
zone-center physical quantities have been found to cluster
increasingly tightly with increasing $J/\hbar\omega$, permitting
an empirical self-trapping curve
\begin{equation}
g_{ST} = 1+\sqrt{ \frac J {\hbar\omega}}
\label{eq:gst}
\end{equation}
to be identified that appears to accurately describe the central
trend of such clusters of data over essentially the entire adiabatic regime.

Similarly, a zone-edge curve
\begin{equation}
g_N = 1 + \sqrt{\frac J {\hbar\omega}} - \left[ 8 \left( \frac J {\hbar\omega} - \frac 1 4 \right) + \left( \frac 2 3 \right) ^8 \right]^{-1/8}
\label{eq:gn}
\end{equation}
 in 1D has been identified that appears to accurately describe
the characteristic changes in the outer energy band that signal the
onset of significant narrowing of the polaron energy band, commencing
the process that develops into the self-trapping transition with
increasing coupling strength.

\begin{figure}[htb]
\begin{center}
\leavevmode
\epsfxsize = 3.6in
\epsffile{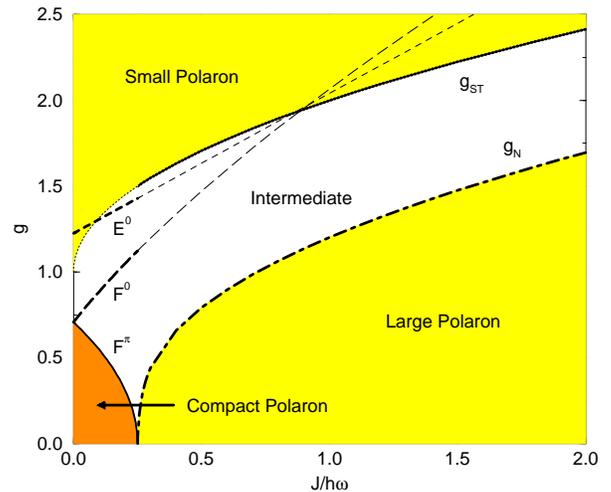}
\end{center}
\caption
{
Polaron phase diagram combining the empirical curves (\ref{eq:ge0}),
(\ref{eq:gf0}), and (\ref{eq:gfpi}) based on the present analysis
employing the Merrifield method for $DJ/\hbar\omega < 1/4$ with the
complementary empirical curves (\ref{eq:gst}) and (\ref{eq:gn}) based
on independent analyses by the Global-Local variational method
for $J/\hbar\omega > 1/4$ in 1D.
}
\label{fig:phasebig}
\end{figure}

It is telling to combine the empirical curves (\ref{eq:gst})
and (\ref{eq:gn}) abstracted from our prior 1D Global-Local analyses
with the complementary curves (\ref{eq:ge0}), (\ref{eq:gf0}),
and (\ref{eq:gfpi}) that follow from our present analysis by
the Merrifield method.
This comparison is presented in Figure~\ref{fig:phasebig}.

It is clear that zone-center properties follow a common pattern of
behavior, relatively indifferent to the crossover from the non-adiabatic
into the adiabatic regime but for the possible role the latter may
play in setting the scale over which the disperse self-trapping
loci of the non-adiabatic regime begin to cluster more tightly
toward the more sharply-defined trend in the adiabatic regime.
This dispersity does not vanish suddenly at the crossover, but is
apparent as well in the dispersity of similar loci found under the
Global-Local method at small $J/\hbar\omega$ still greater than $1/4$.

Further, it is interesting and surely no accident that the
Merrifield zone-center lines (\ref{eq:ge0}) and (\ref{eq:gf0})
intersect, that this intersection falls very nearly upon the
Global-Local zone-center line (\ref{eq:gst}), and that this
cluster of intersections coincides well with the first
appearance of discontinuities in the solutions of the Merrifield method.
It was, in fact, the set of such ``critical points''
($J_c/\hbar\omega , g_c$) ascertained from our sequentially-refined
variational calculations (Merrifield method \cite{Zhao97a}, Toyozawa
method \cite{Zhao97b}, Global-Local method \cite{Brown97b}) that first
led us to identify the simple empirical curve (\ref{eq:gst}) as a
convenient method-independent approximant to the real
physically-meaningful self-trapping line.
The ``critical'' appearance of solutions near such points
($J_c/\hbar\omega , g_c$) is, of course, a methodological artifact,
and we should view the intersections of the several zone-center lines
near $J/\hbar\omega \approx 0.9$ as an artifact as well; the physical
self-trapping line surely bends smoothly through this region,
seamlessly joining the central trend of the non-adiabatic loci with
the more sharply-defined trend that develops at higher adiabaticity.

These zone-center results of the Merrifield method suggest an answer to
one of the more empirical questions left open by our prior Global-Local
analyses.
Though quite accurate over a large range of $J/\hbar\omega$, it has
seemed unlikely that the dependence of the empirical curve $g_{ST}$ on
$\sqrt{J/\hbar\omega}$ should continue unregularized all the way
down to $J/\hbar\omega = 0$.
If we are to continue to regard the physically meaningful $g_{ST}$
as representing a central trend in the inherently disperse set of
self-trapping loci even as $J/\hbar\omega$ vanishes, the present
results suggest that the leading dependence of $g_{ST}$ on $J/\hbar\omega$
should not persist as a square root, but yield to a more pedestrian
linear dependence
\begin{equation}
g_{ST} \sim 1 + a \frac J {\hbar \omega}~, ~~~~~ \frac J {\hbar \omega} < \frac 1 4
\end{equation}
where $a$ is a constant of order unity.

It is likewise clear that the zone-edge loci follow a common
pattern of behavior, albeit one that is exquisitely sensitive to
the crossover from the non-adiabatic into the adiabatic regime.
Although the loci illustrated above and below this crossover are
drawn from different physical properties (Franck-Condon factors from
the Global-Local method are not available and the Merrifield band
energies near the zone edge are not meaningful in and beyond the
crossover regime), they are closely related and reflect the same,
if complementary, underlying physical behavior.

We are led to view the results of the Merrifield method in the
non-adiabatic regime and those of the Global-Local method in the
adiabatic regime as mutually confirming, and describing one
consistent set of self-trapping phenomena at all $J/\hbar\omega$.

In the adiabatic regime, it is quite straightforward to view the
two lines $g_{ST}$ and $g_N$ as dividing the polaron parameter space
into a small polaron regime at strong coupling, a large-polaron regime
at weak coupling, and an intermediate regime occupied by transitional
structures.  In the non-adiabatic regime, it likewise clear that
there is an unambiguous small polaron regime at strong coupling;
moreover, it is noteworthy that the non-adiabatic small polaron
states are continuously deformable into the adiabatic small polaron
states without encountering transition behavior in any basic polaron
property, so that there is no formal distinction to be made between
non-adiabatic and adiabatic small polarons.
A complementary observation can be made about large polarons;
although large polarons as we have thus far characterized them
reside in the adiabatic regime, there is no formal distinction to be
drawn between large polarons at greater or lesser electron-phonon
coupling strengths since these are continuously deformable into
each other without encountering any transition behavior.
The intermediate regime can be defined in similar terms as that in
which transition behavior is found in {\it some} basic polaron
property at every point; for example, although the lines
$g_{ST}$ (zone center) and $g_N$ (zone edge) are only discrete
curves, the domain between them is dense with similar curves
associated with the occurrence of transition behavior at
general $\kappa$ values.

These observations lead us to consider the more darkly-shaded
region of Figure~\ref{fig:phasebig} at the weak-coupling end of
the non-adiabatic regime.
The transition line (\ref{eq:gfpi}) associated with the zone-edge
Franck-Condon factor and the more limited, qualitative information
available in the Merrifield band-edge energies (\ref{eq:gepi})
suggests that this regime is {\it disconnected} from both the
small polaron regime and the large polaron regime in the sense of
continuous deformability as used above.
Provided that this non-adiabatic weak-coupling regime is not
dense with transition loci, which seems quite unlikely, it would
appear that this regime is occupied by a polaron structure that is
neither ``small'' nor ``large'' nor of a transitional nature that
would identify it with the intermediate regime.

Indeed, some very basic polaron properties behave in qualitatively
distinct ways in this insular regime, perhaps foremost being the
polaron radius as given by the polaron Wannier function.
Intuitively, one expects the polaron radius by any definition to
decrease monotonically with increasing electron-phonon coupling, and
this is generally the case in the regimes we have here characterized
as the small polaron regime and the large polaron regime.
Even in the non-adiabatic weak-coupling regime now under discussion,
the radius of the phonon cloud associated with zone-center polarons
has an initial width of order $\sqrt{2J_i/\hbar\omega}$ along
the $i$ axis \cite{Romero98d}, and decreases with increasing
coupling strength.

The polaron Wannier function, however, is a construct of the
{\it entire} polaron energy band, being a superposition of polaron
Bloch states of every $\vec{\kappa}$; as such, it is a localized
state that can be viewed as energy band theory's own answer to the
inverse problem of determining the identity of the localized
quasiparticle whose dynamical properties are manifested in the
polaron energy band.
The real-space width of this state can be gauged in various ways,
among them being a variance measure of the electron density within
the polaron Wannier state.

In present terms, we may construct polaron Wannier states from the
trial Bloch states in the following fashion
\begin{equation}
| \Phi ( \vec{n} ) \rangle = \frac 1 {\sqrt{ N^D}} \sum_{\vec{\kappa}} e^{-i \vec{\kappa} \cdot {\vec{n}}} | \Psi ( \vec{\kappa} ) \rangle ~,
\end{equation}
from which we may construct the electron density profile as
\begin{equation}
\rho_{\vec{r}} = \langle \Phi (0) | a_{\vec{r}}^{\dagger} a_{\vec{r}} | \Phi (0) \rangle.
\end{equation}
Using this density, we may construct a variance tensor
\begin{equation}
\sigma^2_{ij} = \sum_{r_i , r_j} r_i r_j \rho_{\vec{r}}
\end{equation}
in terms of which the spatial variance of the electron density in
the polaron Wannier state may be given in an arbitrary direction.
In the particular case of measurement along the $x$ axis in three
dimensions, for example, this result after summing over the $y$
and $z$ axes takes the form
\begin{equation}
\sigma^2_{xx} = \sum_{r_x,\kappa_x \kappa_x '} \frac {r_x^2} {N_x^2} e^{-i ( \kappa_x - \kappa_x')r_x} \langle \{ \lambda_{\vec{q}}^{(\kappa_x,0,0)} \} | \{ \lambda_{\vec{q}}^{(\kappa_x ',0,0)} \} \rangle ~,
\end{equation}
where
\begin{eqnarray}
\langle \{ \lambda_{\vec{q}}^{(\kappa_x,0,0)} \} && | \{ \lambda_{\vec{q}}^{(\kappa_x ',0,0)} \} \rangle =
\nonumber \\
&& \exp \left[ - \frac 1 {N_xN_yN_z} \sum_{\vec{q}} \left( \frac 1 2 | \lambda_{\vec{q}}^{(\kappa_x,0,0)} | ^2 \right. \right.
\nonumber \\
&& \left. \left. + \frac 1 2 | \lambda_{\vec{q}}^{(\kappa_x',0,0)} | ^2 - \lambda_{\vec{q}}^{(\kappa_x,0,0) *} \lambda_{\vec{q}}^{(\kappa_x',0,0)} \right) \right] ~.
\end{eqnarray}
With further manipulation it can be shown that
\begin{equation}
\sigma^2_{xx} = \frac 1 {N_x} \sum_{\kappa_x} \frac 1 {N_xN_yN_z} \sum_{\vec{q}} \left| \frac {\partial} {\partial \kappa_x} \lambda_{\vec{q}}^{(\kappa_x , 0 , 0)} \right| ^2 ~.
\label{eq:wannsigma}
\end{equation}
The spatial variance of the localized electron density within the
polaron is thus seen to be the average over all phonon modes and over
polaron crystal momenta in the measurement direction of a mean
square measure of the amount of distortion present in the the
phonon amplitudes along the measurement direction.

Without further explicit calculation, this relationship provides a
means of understanding what is distinct in the weak-coupling polaron
behaviors found in the non-adiabatic and adiabatic regimes.
In the adiabatic regime, the weak-coupling polaron band is nearly
identical to that of the free electron in the inner Brillouin zone,
but is strongly flattened in the outer Brillouin zone where the
effects of interaction with the one-phonon continuum are severe.
The phonon amplitudes exhibit strong changes in $\vec{\kappa}$ which,
through (\ref{eq:wannsigma}), are associated with {\it broad} polaron
Wannier states.
With increasing electron-phonon coupling the severity of these
$\vec{\kappa}$-dependent distortions decreases, resulting in the
{\it narrowing} of the polaron Wannier state.
In qualitative terms, this narrowing trend is what is expected of
large polarons.

In the non-adiabatic regime however, quite a different situation is found.
At weak-coupling, the polaron band is nearly identical to that of a
free electron at {\it all} $\vec{\kappa}$, and in the limit of
vanishing coupling is completely undistorted.
The associated phonon amplitudes are not only very small, but
are also very weakly distorted in $\vec{\kappa}$ which, through
(\ref{eq:wannsigma}) implies the complete localization of the
polaron Wannier state on a single site as $g \rightarrow 0$.
Conversely, with increasing electron-phonon coupling, the presence
of the higher-lying one-phonon continuum is felt more strongly at
higher $\vec{\kappa}$, resulting in an enhancement of phonon
amplitudes in the outer zone with which is associated a slight
flattening of the polaron energy band.
This {\it growth} in $\vec{\kappa}$-dependent distortion results
in a {\it broadening} of the polaron Wannier state with increasing
electron-phonon coupling until a transition is made into the small
polaron regime.

Although such {\it compact} polarons 
\footnote{We use the term ``compact'' here to emphasize the complete
spatial localization of polaron Wannier states that distinguishes the
weak-coupling limit of the non-adiabatic regime from that of the
adiabatic regime.
This usage is at least semantically consistent with that of the
theory of nonlinear waves, where particle-like excitations with
finite spatial support have been termed
``compactons'' \cite{Rosenau93,Rosenau94,Rosenau96}.}
are straightforwardly understood, that they are completely localized
in the weak-coupling limit and broaden with increasing electron-phonon
coupling suggests that they be regarded as distinct from both the
large polarons and small polarons that dominate the outer reaches
of the polaron parameter space.
Moreover, that the compact polaron regime appears to be {\it disconnected}
from both the large polaron regime and the small polaron regime by
observable transition behavior suggests that such distinctions may
be important to the clear classification of polaronic systems.

\section{Conclusion}

Our study of the Holstein model has focussed on the basic properties
of observable zero-phonon lines in optical spectra; specifically, the
polaron ground state energies that are principal determinants of the
spectral position of such lines, and the Franck-Condon factors that
are principal determinants of the oscillator strength of such lines.
We have found in these results several properties that facilitate
both the application of these findings to experiments on real bulk
materials and the interpretation of the experimental results in
terms of underlying polaron structure.

First, although quantitative results can be obtained for any
polaron wave vector $\vec{\kappa}$, we have found that at the
extremes of isotropic polaron bands, at the Brillouin zone center and
At its most remote corner, the dependence of the Franck-Condon factor
on real-space dimensionality $D$ and the elementary tunneling
parameter $J$ reduces to the single scaled parameter $DJ$.
This permits a straightforward understanding to be had of how
observations in systems of reduced effective dimensionality, for
example, can be expected to be related to observations in bulk systems.

Second, with some additional care, this general trend in the dependence of
the Franck-Condon factors on dimensionality can be seen to be similar to
that which can be expected in the $\vec{\kappa}$-dependence of the
Franck-Condon factor in a system of fixed dimensionality; for example,
in changing the orientation of the probed wave vector from
$[1,1,1]$ to $[1,1,0]$ to $[1,0,0]$.
Such experimentally-controlled variations in the Franck-Condon
factor (and its associated zero-phonon line) of a fixed system
constitute a signature that can be associated with specific
polaron structure.

Third, beyond such quantitative characteristics we have found that
the Franck-Condon factors at the Brillouin zone center and at the
extreme Brillouin zone edge constitute particularly direct means of
revealing the changes in polaron structure that are associated with
the self-trapping transition.
That is, it appears possible to map out the polaron phase diagram
from surveys utilizing Franck-Condon factors alone.
Through such considerations, here extended for the first time to
the non-adiabatic weak-coupling regime, we have been able to
complete a systematic appraisal of polaron structure spanning
all regimes.
The delineation of transition curves by means of the Franck-Condon
factors has compelled us to distinguish a third kind of characteristic
polaron structure, the {\it compact} polaron, from the more familiar
notions of the {\it small} polaron and {\it large} polaron.
It is suggested that each of these three classes of polaron structure
is separated from the others by an intermediate regime in which
transition behavior is found in some basic polaron properties, but
that this classification is essentially complete.

Thus, suitably discriminating experimental studies of the detailed
behavior of zero-phonon lines would appear to afford versatile,
powerful, and interpretable means of deducing the structure of
polarons in real materials.
Suitably constructed surveys of such structure in a variety of
materials should be capable of mapping such globally-important
features as the polaron self-trapping line, providing experimental
tests of the proposed polaron phase diagram.

\section*{Acknowledgement}

This work was supported in part by the U.S. Department of Energy under
Grant No. DE-FG03-86ER13606.

\appendix

\section{Dimensional scaling}

In some of our illustrations of specific results we take advantage
of a certain scaling property that holds under the Merrifield method
at the Brillouin zone center and at selected points on the Brillouin
zone boundary.
The demonstration of this property utilizes the fact that the
Debye-Waller phases $\Phi_{\mu}^{\vec{\kappa}}$ vanish at the
Brillouin zone center and everywhere on the Brillouin zone boundary,
and the fact that
\begin{equation}
\frac {\partial S_{\mu}^{\vec{\kappa}}} {\partial \lambda_{\vec{q}}^{\vec{\kappa}}} = -4 \frac {\lambda_{\vec{q}}^{\vec{\kappa}}} {N^{D}} S_{\mu}^{\vec{\kappa}} \sin^2 \frac {q_{\mu}} 2,
\end{equation}
as follows from (\ref{eq:Sk}).

Restricting discussion to the zone center ($\vec{\kappa} = \vec{0}$)
and any of the most extreme corners of the Brillouin zone
($\vec{\kappa} = \vec{\pi}$), we find that the fundamental
variational amplitudes can be expressed in the form
\begin{eqnarray}
\lambda^{\vec{0}}_{\vec{q}}  &=& \frac { g \hbar \omega} { \hbar \omega - \frac{N^D} {\lambda_{\vec{q}}^{\vec{0}}} \frac {\partial} {\partial \lambda_{\vec{q}}^{\vec{0}}} \sum_{\mu=1}^D J_{\mu} S^{\vec{0}}_{{\mu}} } ~,\\
\lambda^{\vec{\pi}}_{\vec{q}}  &=& \frac { g \hbar \omega} { \hbar \omega + \frac{N^D} {\lambda_{\vec{q}}^{\vec{\pi}}} \frac {\partial} {\partial \lambda_{\vec{q}}^{\vec{\pi}}} \sum_{\mu=1}^D J_{\mu} S^{\vec{\pi}}_{{\mu}} } ~.
\end{eqnarray}
Now further restricting to the isotropic case, we find
\begin{eqnarray}
\lambda^{\vec{0}}_{\vec{q}}  = \frac { g \hbar \omega} { \hbar \omega - \frac{N^D} {\lambda_{\vec{q}}^{\vec{0}}} \frac {\partial} {\partial \lambda_{\vec{q}}^{\vec{0}}} [ D J S^{\vec{0}} ] } , \\
\lambda^{\vec{\pi}}_{\vec{q}}  = \frac { g \hbar \omega} { \hbar \omega + \frac{N^D} {\lambda_{\vec{q}}^{\vec{\pi}}} \frac {\partial} {\partial \lambda_{\vec{q}}^{\vec{\pi}}} [ D J S^{\vec{\pi}} ] } .
\end{eqnarray}
It is the reduction of the dimension- and $J$-dependence of the
principal quantities to the simple combination $DJ$ that is
responsible for the simplicity of our main results.

\end{document}